\documentclass[amsmath,amssymb,aps,prb,twocolumn,floatfix]{revtex4-2}

\usepackage[caption = false]{subfig}
\usepackage[utf8]{inputenc}
\usepackage{graphicx}
\usepackage{standalone}
\usepackage{dcolumn}
\usepackage{bm}
\usepackage{xcolor}
\usepackage{physics}
\usepackage{amsthm}
\usepackage{enumitem}
\usepackage{mathtools}
\usepackage[colorlinks = true,linkcolor = red,citecolor = magenta]{hyperref}
\usepackage[sort&compress]{natbib}

\DeclareGraphicsExtensions{.pdf,.eps,.png,.jpg,.mps}

\newcommand{\pcsadd}{Center for Theoretical Physics of Complex Systems, Institute for Basic Science (IBS), Daejeon, Korea, 34126}
\newcommand{\ustadd}{Basic Science Program, Korea University of Science and Technology (UST), Daejeon 34113, Republic of Korea}

\makeatletter
\renewcommand*{\fnum@figure}{{\normalfont\bfseries \figurename~\thefigure}}
\renewcommand*{\@caption@fignum@sep}{\textbf{:}}
\makeatother

\newcommand{\FB}{\mathrm{FB}}
\newcommand{\DB}{\mathrm{DB}}
\newcommand{\CLS}{\mathrm{CLS}}

\newcommand{\appropto}{\mathrel{\vcenter{
  \offinterlineskip\halign{\hfil$##$\cr
    \propto\cr\noalign{\kern2pt}\sim\cr\noalign{\kern-2pt}}}}}

\begin{document}

\title{Compact Localized States in Electric Circuit Flatband Lattices}

\author{Carys Chase-Mayoral}
    \email{chasemac@dickinson.edu}
    \affiliation{Department of Physics and Astronomy, Dickinson College, Carlisle, Pennsylvania, 17013, USA}

\author{L.Q. English}
    \email{englishl@dickinson.edu}
    \affiliation{Department of Physics and Astronomy, Dickinson College, Carlisle, Pennsylvania, 17013, USA}

\author{Yeongjun Kim}
    \email{yeongjun.kim.04@gmail.com}
    \affiliation{\pcsadd}
    \affiliation{\ustadd}

\author{Sanghoon Lee}
    \email{scott430@naver.com}
    \affiliation{\pcsadd}
    \affiliation{\ustadd}

\author{Noah Lape}
    \email{lapenoah@dickinson.edu}
    \affiliation{Department of Physics and Astronomy, Dickinson College, Carlisle, Pennsylvania, 17013, USA}

\author{Alexei Andreanov}
    \email{aalexei@ibs.re.kr}
    \affiliation{\pcsadd}
    \affiliation{\ustadd}

\author{P.G. Kevrekidis}
    \email{kevrekid@umass.edu}
    \affiliation{Department of Mathematics and Statistics, University of Massachusetts, Amherst, Massachusetts 01003, USA}

\author{Sergej Flach}
    \email{sflach@ibs.re.kr}
    \affiliation{\pcsadd}
    \affiliation{\ustadd}

\date{\today}

\begin{abstract}
    We generate compact localized states 
    (CLSs) in an electrical diamond lattice, comprised of only capacitors and inductors, via local driving near its flatband frequency.
    We compare experimental results to numerical simulations and find very good agreement. 
    We also examine the stub lattice, which features a flatband of a different class where neighboring compact localized states share lattice sites. 
    We find that local driving, while exciting the lattice at that flatband frequency, is unable to isolate a single compact localized state due to their non-orthogonality.
    Finally, we introduce lattice nonlinearity and showcase the realization of nonlinear compact localized states in the diamond lattice. 
    We induce an instability in the nonlinear CLS when it is shifted into resonance with a dispersive (optical) band.
    Our findings pave the way of applying flatband physics to complex electric circuit dynamics.
\end{abstract}

\maketitle

\section{Introduction} 
\label{introduction}

Flatbands (FBs) arise as completely degenerate energy bands in certain tight-binding lattices with macroscopic degeneracy~\cite{leykam2013flat, leykam2018artificial}.
FB signals zero group velocity, suppressing transmission, and producing compact localized states (CLSs), i.e., eigenstates trapped in a strictly finite number of sites.
The extreme sensitivity of FBs to perturbations leads to the emergence of diverse and intriguing phases as it lifts the macroscopic degeneracy, including ferromagnetism~\cite{derzhko2015strongly, lieb1989two, mielke1991ferromagnetism, tasaki1992ferromagnetism, mielke1999stability}, superfluidity~\cite{peotta2015superfluidity, julku2016geometric, tovmasyan2018preformed, mondaini2018pairing, aoki2020theoretical},
localization-delocalization transition~\cite{goda2006inverse,cadez2021metal, kim2022flat, lee2023critical, lee2023critical2}, many-body flatband localization~\cite{kuno2020flat_qs, danieli2020many, vakulchyk2021heat, danieli2022many, tilleke2020nearest}, symmetry-breaking transitions~\cite{vicencio2013discrete,nguyen2018symmetry},
and compact discrete breathers~\cite{danieli2021nonlinear, danieli2018compact}, among others~\cite{derzhko2015strongly,leykam2018artificial,leykam2018perspective,PRR2043426}.

With the above, the experimental realization of artificial FB lattices becomes a priority. 
The challenge is in fine-tuning to preserve the CLS over a long time.
Several experimental attempts, typically over a short time, lacked crucial relative phase information for the CLS and sufficient spatial resolution.
Examples are photonic lattices~\cite{nakata2012observation, mukherjee2015observation,kajiwara2016observation, vicencio2015observation, nguyen2018symmetry, ma2020direct}, cold atoms~\cite{taie2015coherent, ozawa2017interaction}, polariton condensates~\cite{baboux2016bosonic, masumoto2012exciton}, 
electrical circuits~\cite{zhang2023non, wang2022observationof,wang2019highly, zhou2023observation} and topological material~\cite{kang2020topological}, as well as magnonic~\cite{tacchi2023experimental} crystal lattices.
Electrical circuits offer a particularly promising approach for in-depth exploration of FBs and CLS.
They provide flexibility in various lattice geometries, feasibility of fine-tuning lattice parameters, and precise experimental control and measurement.

In this paper, we experimentally construct and characterize one-dimensional FB electrical lattices with discrete circuit elements.
We excite FB CLS through local sinusoidal driving at the FB frequency, reporting results of two lattice structures: diamond and stub lattices.
The diamond lattice contains orthogonal CLSs and exhibits CLS resonant modes when locally driven at the FB.
Finally, we impart nonlinearity by replacing the capacitors with varactor diodes characterized by a voltage-dependent capacitance.
Our findings reveal the continuation of CLSs into the highly nonlinear regime in the diamond lattice and the onset of nonlinear CLS instability when it becomes resonant with a dispersive band.
On the other hand, a stub lattice features non-orthogonal CLSs and shows exponentially localized resonant modes by overlapping CLSs.

\section{The model -- Diamond chain}
\label{the_model_diamond}

\begin{figure}
    \includegraphics[width=3.45in]{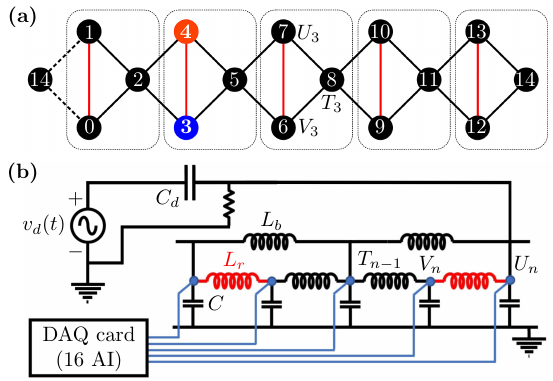}
    \caption{
        (a) Tight binding representation of the diamond lattice, with grey boxes denoting unit cells and red/blue circles representing CLS with opposing amplitudes.
        (b) The lattice is driven at one site via a driving capacitor, \(C_d\), by a sinusoidal voltage signal, \(v_d(t)\), from a signal generator.
    }
    \label{fig:setup}
\end{figure}

In the electrical-lattice context, vertices and edges of a tight-binding lattice represent capacitors and inductors, respectively.
The lattices are examples of electrical transmission lines with nontrivial geometry, as depicted in Fig.~\ref{fig:setup}~(a) and (b), where a diamond lattice is shown with two different hopping values denoted by black and red lines.
The capacitance of each node is \(C\), and the lattice incorporates inductors with different inductance, \(L_b\) and \(L_r\), shown as black and red, respectively.
The inductors are assumed to have an effective serial resistance (ESR) \(R\), while the capacitors are treated ideally.

The voltages at the three nodes of the \(n\)-th unit cell are denoted as \(T_n, U_n, V_n\).
We drive the lattice at \(U_{m}\) in the \(m\)-th unit cell with a driving capacitor of capacitance \(C_d \ll C\).
From Kirchhoff's current law at each node, the equations of motion for the voltages in the \(n\)-th unit cell at the linear level take the form:
\begin{align}
    \label{eq:diamond_undriven}
    \ddot{T}_n +\beta \dot{T}_n &= -\omega^2_b\left[4T_n - U_n -U_{n+1} - V_n -V_{n+1}\right], \notag \\
    \ddot{U}_n \!+\beta \dot{U}_n &= -\omega^2_b\left[\left(2 + \alpha\right) U_n -\alpha V_n - T_n-T_{n-1}\right], \\
    \ddot{V}_n +\beta \dot{V}_n &= -\omega^2_b\left[\left(2 + \alpha\right) V_n -\alpha U_n - T_n-T_{n-1}\right] \notag.
\end{align}
At the driven site \(U_m\), Eq.~\eqref{eq:diamond_undriven} gets a correction factor \(1/(1+\gamma)\) and an additive driving force \(A\sin(\omega_d t)\),
\begin{align}
    \label{eq:diamond_driven}
    \ddot{U}_m \!+\! \frac{\beta\dot{U}_m}{1\!+\!\gamma} \!&= \! \frac{-\omega^2_b}{1\!+\!\gamma}\left(\!\left(2 \!+\! \alpha\right) U_m \!-\! \alpha V_m \!-\! T_m \!-\! T_{m-1} \vphantom{\sum}\!\right) \\
    \!&+\! A\sin(\omega_d t). \notag 
\end{align}
\(\gamma= C_d/C\), where \(C_{d} \ll C\),  is an impurity artifact that appears as a result of driving which can be minimized to within the experimental tolerance of \(\omega_b^2\).
Note that \(\omega_b^2 \!\!=\!\! 1/(L_bC)\) and \(A= v_d\omega_d^2\gamma(1+\gamma)^{-1}\).
\(\alpha=L_b/L_r\) tunes the flatband, and \(\beta = R/L\) accounts for dissipation.
Then, the equation of motion is approximately written as follows,
\begin{align}
    \label{eq:diamond_schrodinger_eq}
    \left(\frac{d^2}{dt^2} + \beta \frac{d}{dt} \right) \ket{\psi(t)} = H \ket{\psi(t)} + \ket{\mathbf{F}(t)}.
\end{align}
This equation describes wave dynamics on a diamond lattice.
Here, the wave, \(\ket{\psi(t)} = \sum_{n, X} X_n\ket{X_n}\), is the voltage at each node, where $\ket{X_n}$ is real-space lattice site basis at sublattice $X \in \{T, U, V\}$, and \(n \in \mathbb{Z}\).
The matrix \(H\) represents the coupling of the diamond lattice, given from the right side of Eq.~\eqref{eq:diamond_undriven}.
\(\ket{\mathbf{F}(t)}\) is the driving term, yielding the last term of Eq.~\eqref{eq:diamond_driven}.
Unlike the hoppings in tight-binding diamond lattices, inductors also add to the ``onsite potential'' (diagonal elements of \(H\)), which in turn guarantees \(\omega^2\) is positive and breaks the lattice bipartitness and chirality, but not the flatband~\cite{calugaru2022general}.
In the absence of driving, the Bloch waveform ansatz, \(U_n = U(k)\exp\left[i(\omega t - kn)\right]\) (and similarly for \(V_n, T_n\)) solves the eigenvalue problem associated with Eq.~\eqref{eq:diamond_undriven}.
It results in three bands - an optical band, a gapped flatband, and lower frequency acoustic band which extends down to zero frequencies:
\begin{align}
    \label{eq:diamond_eigenvalues}
    \omega_{\FB}^{2} = 2\omega^{2}_b(\alpha + 1),
    \hspace{0.45em} \omega_{\DB}^2 = \omega^2_b(3\pm \! \sqrt{4\cos(k) + 5}).
\end{align} 
Solving the dissipative case yields \(s^2 \!-\! i\beta s \!=\! \omega^2_{\FB/\DB}\),
\begin{align}
    \label{eq:diamond_eigenvalues_complex}
    s_{\FB/\DB} = i\frac{\beta}{2} \pm \sqrt{ - \frac{\beta^2}{4} + \omega^2_{\FB/\DB}},
\end{align}
with the dissipation time \(\tau \!=\! 2/\beta\) and the slightly shifted \(\omega_{\FB/\DB}\) (\(<\!1\%\) in our experiment) from Eq.~\eqref{eq:diamond_eigenvalues}.
We assume only underdamped frequencies making the square root part always real, i.e. \(\omega^2_{\FB/\DB} > \beta^2/4\).
Hence, we set \(s_{\FB/\DB} \approx \omega_{\FB/\DB}\) throughout the paper.

While the Bloch eigenvectors at FB are spatially extended, the degeneracy of the FB allows CLS eigenstates.
For the diamond lattice, the CLS is given by \(U_n =\delta_{nm}\), \(V_n = -U_n\) and \(T_n = 0\) for any unit cell \(n\) forming an orthogonal basis in the FB subspace, as we will see below in further detail in Fig.~\ref{fig:diamond_result}.
A similar analysis is performed on the stub lattice (Fig.~\ref{fig:stub}~(b)), and is discussed in Sec.~\ref{failure_of_cls_excitation_stub}.

\section{Response to local perturbation}
\label{response_to_local_perturbation}

We analyze the impact of local driving on the flat-band analytically in this section, and we apply this result to the diamond lattice and stub lattice in the following sections.

\begin{figure*}
    \includegraphics[width=\textwidth]{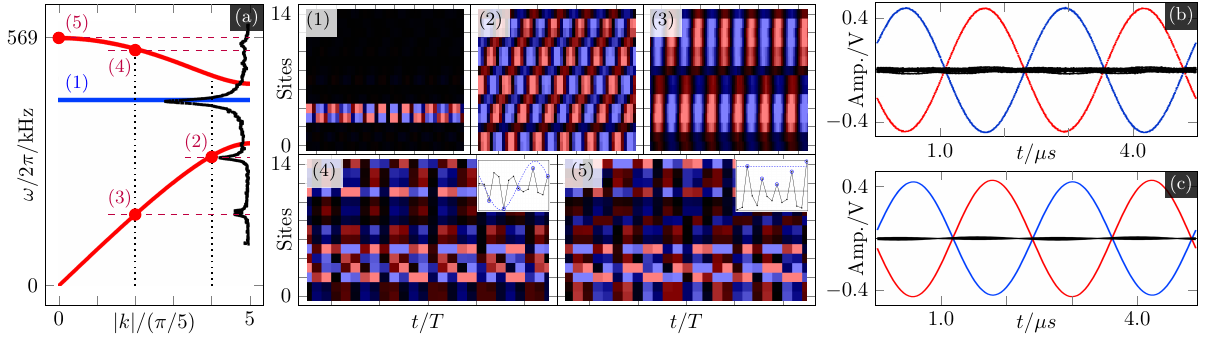}
    \caption{
        (a) The lattice response to local driving (at node 4) as a function of driving frequency (exciting wavenumbers \(\pm k\)). 
        The system features two dispersive bands - one acoustic, one optical branch, and one flatband. 
        In panels (1) -- (5), the driving frequency \(f_d\) is chosen according to the labels in (a).
        The time axis is in units of \(T = 1/f_d\); red, blue and black indicate positive, negative and zero voltages, respectively.
        Driving near the flatband yields a CLS (1), driving in the acoustic branch yields a spatially extended response (2), (3).
        Two examples in the optical branch are shown in (4), (5), where inset shows the spatial voltage profile (consistent with the corresponding \(k\)-value of the Bloch eigenfunction) at a moment in time.
        (b) Response at flatband corresponding to the highest peak in (a), \(f_d = 401\text{ kHz}\), see panel (1). 
        Blue and red colored lines correspond to the experimental CLS sites 3, 4.
        (c) Simulation result of Eq.~\eqref{eq:diamond_undriven} with experimental parameters at \(429\) kHz.
    }
    \label{fig:diamond_result} 
\end{figure*}

To analyze the impact of local driving on the flatband, we employ the transfer function method~\cite{economou2006green}, a powerful tool to understand the dynamic behavior of systems described by linear differential equations.
In our case, the transfer function operator \(G(\omega)\) is defined in the following:
\begin{align}
    \left(-\omega^2 + i\beta\omega
    -H\right) G(\omega) = I
\end{align}
The transfer function operator $G(\omega)$ encodes all information on the response to local sinusoidal driving.
With zero initial condition, the response to the ideal sinusoidal driving is then given as:
\begin{gather}
    \label{eq:psi_computed_using_gf}
    \ket{{\psi}(\omega)} = G(\omega)\ket{\mathbf{F}(\omega)},
\end{gather}
The transfer function operator \(G(\omega)\) can be easily calculated in the eigenbasis of \(H\):
\begin{align}
    \label{eq:green_diagonal_form_separate}
    G(\omega) &= G_{\FB}(\omega) + G_{\DB}(\omega) \notag \\
    &= \frac{P_{\FB}}{i\beta\omega \!-\! \omega^2 \!+\! \omega^2_{\FB}} +\! \sum_{j, k}\frac{\ketbra{\psi_j(k)}{\psi_j(k)}}{i\beta\omega \!-\! \omega^2 \!+\! \omega^2_j(k)}.
\end{align}
Note that we treat the frequency responses of the flat band and dispersive bands seperately, to isolate the flatband response, where \(j \in \DB\) is the band index for the dispersive bands. 
If we assume ideal local sinusoidal driving at \(Y_m\), we can write as
\begin{align}
    \label{eq:flatband_response}
    \ket{\mathbf{F}(\omega)} = \frac{A}{2}(\delta(\omega - \omega_d) + \delta(\omega + \omega_d))\ket{Y_m}.
\end{align}
We assume local driving at \(Y_m\), where \(Y=T,U,V\) is a sublattice index.
We neglect the dispersive term \(G_{\DB}\), which is reasonable when \(\omega_d \approx \omega_{\FB}\) and the dispersive bands are sufficiently far from the flat bands compared to the width of resonance peaks.
We use Eqs.~\eqref{eq:psi_computed_using_gf}, \eqref{eq:green_diagonal_form_separate},\eqref{eq:flatband_response} and obtain the spatial profile
\begin{align}
    \label{eq:flatband_response_to_local_driving}
    \braket{X_n}{\psi(\omega)} \approx \mel{X_n}{G_{\FB}}{Y_m} \propto \mel{X_n}{P_{\FB}}{Y_m}.
\end{align}
We express the projector in terms of the CLS basis,
\begin{align}
    \label{eq:projector_nonorthogonal}
    P_{\FB} = \!\! \sum_{i,j \in \mathbb{Z}} \!\! \left[S^{-1}\right]_{ij} \! \ketbra{\CLS_i}{\CLS_j},
\end{align}
where \(i, j\) are lattice sites, and \(S\) is the \emph{overlap matrix} with elements defined as \(S_{ij} \!=\! \braket{\CLS_i}{\CLS_j}\).
For flatbands supporting orthogonal CLSs, \(S_{ij} = \delta_{ij}\) (Eq.~\eqref{eq:projector_nonorthogonal}), the projector \(P_{\FB}\) is expressed simply as
\begin{align}
    \label{eq:orthogonal_cls_projector}
    P_{\FB} = \sum_{i\in\mathbb{Z}} \ketbra{\CLS_i}{\CLS_i},
\end{align}
resulting in a \emph{compact} projector~\cite{sathe2021compactly}. 
The term \emph{compact} implies the existence of an integer \(l \leq |n-m|\), such that the matrix elements satisfy \(\langle{X_m | P_{\FB} | Y_n}\rangle = 0\), where \(X, Y \!\in\! \{T, U, V\}\).
For instance, for the diamond lattice (Sec.~\ref{the_model_diamond}), the CLS at the \(i\)-th unit cell is represented as
\begin{gather}
    \label{eq:diamond_cls}
    \ket{\CLS_{i}} = \left(\ket{U_{i}} - \ket{V_{i}}\right)/\sqrt{2}.
\end{gather}
Then, the real-space representation of the projector is obtained from Eq.~\eqref{eq:flatband_response_to_local_driving}, and the CLS response of the diamond lattice at \(U_n\) is obtained as
\begin{gather}
    \label{eq:diamond_CLS_response}
    |\braket{U_n}{\psi(\omega)}| = \frac{1}{2}\frac{\delta_{mn}\gamma(1+\gamma)^{-1}v_d \omega^2_d}
    {\sqrt{(\omega_{FB}^2 - \omega_d^2)^2 + \beta^2\omega_d^2}},
\end{gather}
which is identical for the \(V_n\) sites.
For the parameters in the experimental setup, we have \(\gamma = 0.015\), \(v_d = 1 \textrm{ V}\), \(\omega_d = \omega_{\FB} = 2\pi \times 429 \textrm{ kHz}\), and \(\beta = R/L_b = 49356\textrm{/sec}\).
The maximum CLS response is given by \(0.4\textrm{ V}\), which is in excellent agreement with the experiment and simulation within 10~\% (Fig.~\ref{fig:diamond_result}~(b) and (c)).

\section{Experimental and numerical results -- Diamond}
\label{experimental_and_numerical_results_diamond}

We construct an electrical-circuit diamond lattice of \(N = 5\) unit cells, with periodic boundary conditions.
The lattice includes \(3N = 15\) capacitors, each with a capacitance of \(C=1 \pm 0.01~\mathrm{nF}\); the driving capacitor of \(C_d=15~\mathrm{pF}\) yields \(\gamma = 0.015\).
Two inductors are employed with \(L_b=466~\mu\mathrm{H}\) and \(L_r=674~\mu\mathrm{H}\), within a 1\% tolerance.
The main sources of inductor dissipation include ferrite cores and coil-wire resistance, diminishing the quality factor \(Q=\omega L/R_\mathrm{eff}\).
The \(Q\) of an inductor remains essentially constant while the effective serial resistance (ESR) varies according to the resonant frequency.

With \(Q=55\) at \(232~\mathrm{kHz}\), we estimate ESR \(R_\mathrm{eff} \approx 23~\Omega\) for the \(L_b\) inductor at \(f_{\mathrm{FB}}\)~\cite{bourns}.
A \(10~\mathrm{k}\Omega\) resistor is placed in parallel with the lattice to suppress a DC voltage component.
We then obtain the band structure using Eq.~\eqref{eq:diamond_eigenvalues} (red (dispersive)  and blue (FB) lines in Fig.~\ref{fig:diamond_result}~(a)).
The flatband frequency \(f_{\mathrm{FB}} = 429~\mathrm{kHz}\) is in the spectral gap between the two dispersive bands.
The flatness of this band has been ascertained experimentally by measuring its frequency using both a spatially uniform and a staggered driver.

To experimentally probe the flatband and its CLS, we supply energy locally via a sinusoidal voltage input from a signal generator (Agilent 33220A function/sweep generator), see Fig.~\ref{fig:setup}.
The measurement results are displayed in Fig.~\ref{fig:diamond_result}. 
We also monitor the response voltage at all lattice sites, which corresponds to \((U_0, V_0, T_0, \ldots, U_{4}, V_4, T_4)\), simultaneously with a 16-channel data acquisition system (NI PXI-1033 with NI 6133 cards) at a \(2.5~\mathrm{MHz}\) sampling rate. 
The driving voltage is injected at site 3 (\(U_1\)).
While we show a continuous band structure, there is only \(N=5\) resonance modes per band, hence \(\lceil N/2 \rceil=3\) peaks per band due to two-fold degeneracy.

Let us turn to the impact of local driving frequency \(f_d\).
We sweep \(200\) -- \(600~\mathrm{kHz}\) frequency range within \(25~\mathrm{ms}\) using a function generator.
The steady-state amplitude responses at site 4 (\(V_1\)) were measured with an oscilloscope (no DAQ card), shown as the black trace in Fig.~\ref{fig:diamond_result}~(a).
The flatband frequency prediction is accurately matched.
This resonance peak strength depends on dissipation, driving voltage, and amplitude of resonant eigenvector at \(U_m\).
We observe the largest peak reaching \(0.4~\mathrm{V}\) at \(429~\mathrm{kHz}\).
Two other prominent modes in the acoustic branch are also visible at \(k=2\pi/5, 4\pi/5\).
We now tune the function generator to the frequencies of the observed resonance peaks.
Fig.~\ref{fig:diamond_result}~(1) -- (5) display spatial patterns at various drive frequencies once they reach a steady state.

At \(f_d = 429~\mathrm{kHz}\), corresponding to the largest resonance peak on the flatband, the associated CLS is predicted to sit at \(U_n, V_n\) of a single unit cell, with their excitations out of phase.
Figs.~\ref{fig:diamond_result}~(b) and (c) compare experiment and numerics, respectively.
In Fig.~\ref{fig:diamond_result}~(b), the red trace depicts the response of the driven site, whereas the blue depicts that for the other CLS site.
The traces show voltage-time profiles for all 15 sites within the time interval \((0, 2/f_d)\).
We observe precise out-of-phase behavior, causing destructive interference at neighboring bottleneck sites \(T_n, T_{n-1}\).
However, small leakage to the rest of the sites is evident due to experimental imperfections and dissipation, which broadens the dispersive resonance peaks
-- yet, this does not adversely affect the CLS (even in the nonlinear regime, as we will see), as its frequency is detuned from any other eigenmodes.
Note that driving at a site in the CLS is essential for its generation.

In Fig.~\ref{fig:diamond_result}~(c), the corresponding numerical simulation demonstrates excellent agreement with the experiment.
In the simulation, for \(\beta = 0\), all sites other than the CLS sites approach to near-zero voltages, 
resulting in a true CLS localized on the two CLS sites, oscillating at the flatband frequency of \(429~\mathrm{kHz}\). 
Note that the experimental frequency is shifted down slightly to \(401~\mathrm{kHz}\) due to small parasitic capacitances associated with the measurement apparatus (ribbon cables and DAQ board).

\section{Experimental and numerical results -- Stub chain}
\label{failure_of_cls_excitation_stub}

\begin{figure*}
    \includegraphics[width=\textwidth]{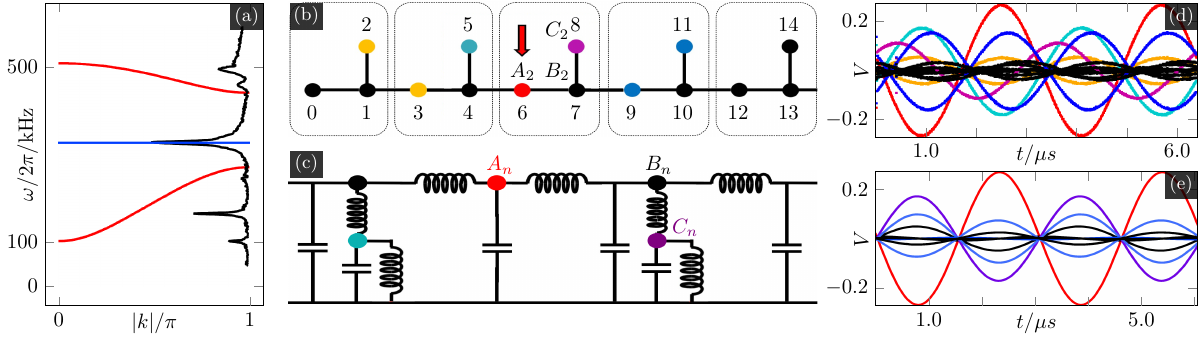}
    \caption{
        Spectrum of solutions for the stub lattice, according to Eq.~\eqref{eq:stubdisp} - the red curves indicate dispersive bands and blue curve is the flat band.
        The experimental spectrum is displayed by the black trace along the right vertical axis, obtained by frequency-sweeping the local driver at a CLS site.
        Panels (b) and (c) show a stub lattice schematic, and the electrical circuit implementation, respectively. 
        In-stub tight-binding diagram, (+) and (-) at sites \(5\), \(8\) and the site \(6\), represent a CLS.
        The colors of the curves in panel (d) and (e) pertain to the respective ones used in panel (b).
        Panel (d) shows the experimental result of local driving at site \(6\), where the trace color assignment and driving location (arrow) is given in (b). 
        Panel (e) displays the corresponding numerical result.
        The plot of site \(5\) (cyan) is hidden behind the plot of site \(8\) (purple).
        This is also the case for sites \(2\) and \(3\) which are hidden behind the plot of sites \(9\) and \(11\).
    }
    \label{fig:stub} 
\end{figure*}

We conduct a similar analysis on a stub lattice.
In Fig.~\ref{fig:stub}~(b) and (c), we illustrate the circuit representation and schematic of the stub lattice.
In the stub case, only a single type of inductor is required  (we use \(L_{b}\)).
The equations of motion for the stub lattice are
\begin{align}
    \label{eq:stub_undriven}
    \ddot{A}_n +\beta \dot{A}_n &= -\omega^2_b\left[2A_n -B_{n-1} - B_n \right], \notag \\
    \ddot{B}_n +\beta \dot{B}_n &= -\omega^2_b\left[3B_n - C_n - A_{n} - A_{n+1}\right], \\
    \ddot{C}_n +\beta \dot{C}_n &= -\omega^2_b\left[2C_n - B_{n}\right].
\end{align}
The driven site is \(A_m\) and is modified similarly to Eq.~\eqref{eq:diamond_driven}.
The eigenfrequencies for the stub lattice for \(\beta = 0\) are computed as
\begin{gather}
    \label{eq:stubdisp}
    s^2 =
    \begin{cases}
        \omega^2_{\FB} = 2\omega^2_b, \\
        \omega^2_{\DB} = (\omega^2_b/2)(5 \pm\! \sqrt{13 + 8\cos(k)}).
    \end{cases}
\end{gather}
Similar to the diamond chain the onsite potentials break chirality (but, again, not the flatband~\cite{calugaru2022general}), and in addition here also make both dispersive bands optical-like, with a finite gap to zero frequencies.
The blue and red traces in Fig.~\ref{fig:stub}~(a) plot Eq.~\eqref{eq:stubdisp}; the black trace along the right vertical axis depicts the experimental spectrum obtained by sweeping the frequency of the driver at a CLS site.
Note that the largest experimental response at this CLS site is registered at the predicted flat-band frequency.
The zone-center acoustic mode occurs at a nonzero frequency and is registered in the spectrum.
At \(\omega_{\FB}\), the CLS labeled with \(i\) occupies the \(i\)-th and \((i+1)\)-th unit cells:
\begin{gather}
    \label{eq:stub_cls}
    \vert \CLS_i \rangle = \frac{1}{\sqrt{3}}\left(\ket{C_{i}} + \ket{C_{i+1}} - \ket{A_{i+1}}\right).
\end{gather}
As a result of the overlap between the closest CLSs, they form a nonorthogonal basis.
Then, the impact of local driving on the stub flatband exhibits distinct behavior compared to the diamond case, leading to \(S_{ij} \neq 0\) for \(i\neq j\).
Given that the stub CLSs only overlap with their nearest CLSs, as defined by Eq.~\eqref{eq:stub_cls}), the overlap matrix \(S_{ij}\) in Eq.~\eqref{eq:projector_nonorthogonal}) is determined as
\begin{align}
    \label{eq:overlap_matrix_nearest_neighbor}
    S_{ij} = \braket{\CLS_i}{\CLS_j} = \delta_{ij} + \sigma \delta_{i \pm 1, j},
\end{align}
where the overlap between neighboring CLSs is denoted as \(\sigma\).
\(S\) is a tridiagonal matrix with translational symmetry, thus its inverse can be readily obtained in the momentum basis,
\begin{align}
    \left[S^{-1}\right]_{ij} \!=\! \frac{1}{2\pi}\!\int^{\pi}_{-\pi} \!\!\!\! dk\frac{\exp(ik|i-j|)}{-1 - 2\sigma \cos (k)} \propto e^{-|i-j|/\xi}.
\end{align}
The integration is solved in the complex plane using Cauchy's integral formula with substitution,
\(\omega \!=\! \exp(ik)\) and \( d\omega \!=\! i\omega dk\)~\cite{sheng2007introduction}.
Here, a characteristic localization length \(\xi\) is obtained as,
\begin{gather}
    \frac{1}{\xi} = \ln \abs{\frac{2\sigma}{-1 + \sqrt{1-  4\sigma^2}}}.
\end{gather}
Therefore, local driving induces exponential localization around the driven site, not an excited CLS mode.
We have \(\sigma = 1/3\) and thus \(\xi \approx 1.03\) for the CLS in Eq.~\eqref{eq:stub_cls}.

To verify the theoretical prediction, we introduce a driver at site 6 (indicated in red arrow) as depicted in Fig.~\ref{fig:stub}~(b), it is expected to induce partial excitation in both CLSs which share the site 6.
This situation experimentally shown in Fig.~\ref{fig:stub}~(d), using a driver frequency of \(312~\mathrm{kHz}\) with a driving amplitude of \(11~\mathrm{V}\).
The corresponding result is shown in the numerical simulation of Fig.~\ref{fig:stub}~(e).
The voltage-time profiles of all 15 sites are presented for two periods.
When site 6 (red) is driven, neighboring CLSs also undergo excitation (depicted in yellow and blue).
This phenomenon arises due to the shared sites 5 and 8 with adjacent CLSs.
It is important to note that slight inhomogeneities result in uneven excitation amplitudes between the two neighboring CLSs.

\begin{figure*}[tbh]
    \centering
    \includegraphics[width=\textwidth]{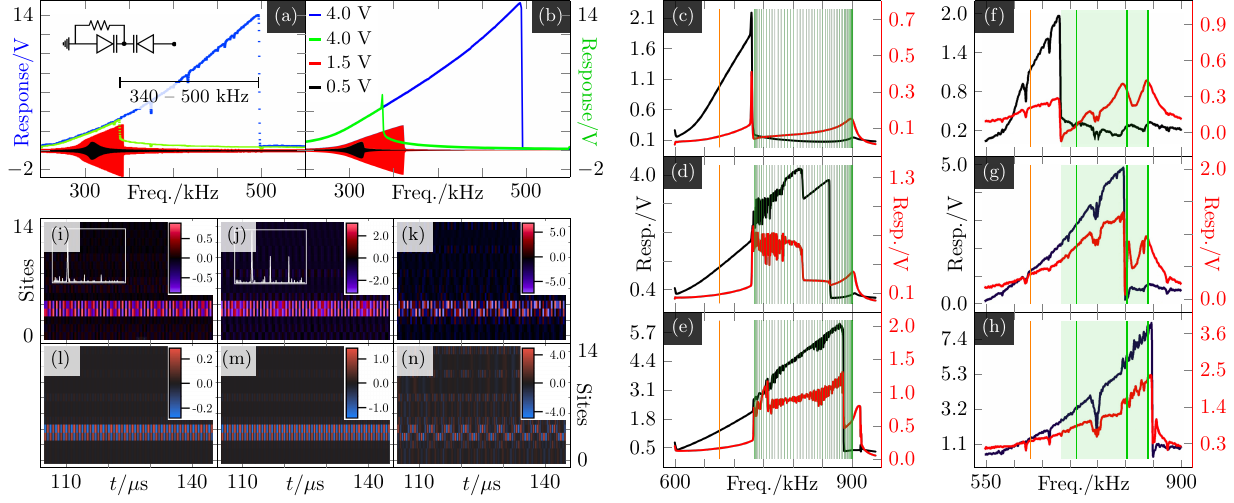}
    \caption{
        The nonlinear system is achieved by (a) replacing the capacitors with the element shown in the inset.
        (b) corresponds to the numerical simulation of (a) to obtain the coupling \(g\) in Eq.~\eqref{eq:nonlincoup}.
        A hysteresis window opens up at the highest amplitude (blue, green).
        (c) -- (e) show the resonance curve obtained from numerical simulations for \(N = 100\) unit cells at \(v_{d} = 0.5, 1.0, 2.0\) from top to bottom. 
        (f) -- (h) show the resonance curves obtained experimentally for local driving.
        The black (red) trace is the response of site 3 (2).
        (i) -- (k) display the full spatio-temporal pattern for \(N = 5\) unit cells from the experiment.
        At low and intermediate amplitudes outside the optical band, the CLS remains stable: (i) \(v_d=0.5\text{ V}\) with \(f_d= 578\text{ kHz}\) and (j) \(v_d=2.0\text{ V}\) with \(f_d= 650\text{ kHz}\).
        The insets show the FFT of response at site \(3\), indicating the emergence of higher harmonics.
        At high amplitudes, such as (k) \(v_d=4.0\text{ V}\) and \(f_d= 735\text{ kHz}\), CLS instability occurs when resonating with an optical band  mode.
        (l) -- (n) depict the full spatio-temporal pattern from the numerical simulations.
        At low and intermediate amplitudes outside the optical band, we see stable CLS: (l) \(v_{d} = 0.5\text{ V}\) with \(f_{d} = 638\text{ kHz}\) and (m) \(v_{d} = 2.0\text{ V}\) with \(f_{d} = 680\text{ kHz}\).
        At high amplitudes, such as (n) \(v_{d} = 4.0\text{ V}\) and \(f_{d} = 797\text{ kHz}\), CLS instability is once again observed.
    }
    \label{fig:nonlinear}
\end{figure*}

\section{Nonlinear compact localized states}
\label{excitation_of_nonlinear_cls_mode}

We extend our studies and investigate the impact of nonlinearity on the CLS in the diamond chain both experimentally and numerically.
It has been predicted that linear homogeneous CLSs (absolute values of all nonzero amplitudes are equal) can be extended into the nonlinear regime as families of compact periodic orbits or compact discrete breathers in the presence of a suitable symmetric nonlinearity~\cite{danieli2018compact}. 
In order to create a symmetric hard-type nonlinearity, we substitute the capacitors with \textit{varactors} -- diode pairs oriented in opposite directions~\cite{fukushima1980envelope}.
The varactor configuration is illustrated in the inset of Fig.~\ref{fig:nonlinear}~(a).
The additional ground-connected resistor (\(100~\mathrm{k}\Omega\)) is needed to prevent a DC charge buildup.
It breaks the symmetry, but the effect is weak for the chosen large resistance value due to the small current actually flowing through it.
The diodes ensure a symmetric nonlinear current-voltage characteristics. 
Consequently, the term \(\omega^2_b\) in Eq.~\eqref{eq:diamond_undriven}, becomes a nonlinear symmetric function of the voltage \(U_n\), at sites \(U_n\) (and similarly for \(V_n, T_n\)). 

In order to experimentally demonstrate nonlinear CLS, we employ these diodes and a \(680~\mu\text{H}\) inductor to build an RF-resonator.
We drive it using a sweep generator and a linear capacitor, recording the resulting resonance curves.
In Fig.~\ref{fig:nonlinear}~(a), a low driving amplitude yields a symmetric peak (black).
As we increase the driving amplitude (red), the curve shifts to higher frequencies.
At \(v_d = 4\text{ V}\), a significant bistability window emerges, (340 -- 500~{kHz}), with hysteresis evident in the up and down sweeps (blue and green, depicting only peak-to-peak amplitude for visual clarity).

We model the nonlinearity with the following ansatz:
\begin{gather}
    \label{eq:nonlincoup}
    \omega^{2}_{b}(U_n) = \omega^{2}_{b0}\left[1 + \ln(1 + gU^{2}_{n})\right],
\end{gather}
where \(g =0.1\) characterizes the hard-type nonlinearity, and \(\omega_{b0} =  \omega_b(0)\).
This particular choice of nonlinearity provides a good fit to our experimental data at strong driver voltage, see the blue curves in Fig.~\ref{fig:nonlinear}~(a), and Fig.~\ref{fig:nonlinear}~(g),~(h).
At weak voltage, \(gU_{n}^{2} \! \ll \! 1\), the model simplifies to a  quadratic form, \(\omega^{2}_{b} \to \omega^{2}_{b0}(1 + gU^{2}_{n})\), capturing the essence of nonlinearity at lower voltages.
In Appendix.~\ref{stability_of_undriven_nonlinear_cls}, we conduct a stability analysis following Ref.~\cite{danieli2018compact} for the undriven nonlinear diamond lattice, for quadratic, and logarithmic nonlinearities.

In the nonlinear diamond lattice, we simulate frequency sweeps starting near the flat band -- see Fig.~\ref{fig:nonlinear}~(c) -- (e).
When reaching the optical band-edge for this lattice of 100 unit cells, 735 kHz, the signal either drops abruptly to zero or (at higher amplitudes) continues with additional noise.
The black (red) curves represent the response at a CLS (non-CLS) site.
Notably, panel (c) is essentially unchanged over a range of driver amplitudes (\(A\) up to \(0.8\text{ V}\)), and upon entering the optical band (\(A \geq 0.9\text{ V}\)), the response at the adjacent non-CLS site jumps up discontinuously.

In the experiment, panels (f) -- (h), we observe similar behavior in a smaller lattice (\(N=5\)).
These panels show the response again for different driving amplitudes (top \(1\text{ V}\), middle \(2.25\text{ V}\), bottom \(2.75\text{ V}\)). 
The two optical modes show up prominently in the red curve.
More importantly, when the driving frequency reaches the first of these modes at \(800\text{ kHz}\), it either drops or continues with enhanced noise.

Fig.~\ref{fig:nonlinear}~(i) -- (n) display  the spatio-temporal profiles at stationary resonant states at multiple driving amplitudes, both experimentally and numerically.
We start in the nearly linear regime at \(578^{*}\text{ kHz}\) (equivalent to \(630\text{ kHz}\), shift due to DAQ/ribbon cable), and raise the frequency to \(650^{*}\text{ kHz}\) (\(708\text{ kHz}\)) with a stronger driver.
This procedure does not disrupt the CLS.
When extending deeper into the nonlinear regime, the emergence of harmonics in the CLS spectrum (see insets) underscores the nonlinear nature of the stationary CLS.
Lowering the amplitude while maintaining \(650\text{ kHz}\) destroys the CLS due to the nonlinearity-induced bistability and hysteresis.

Exploiting the strong nonlinearity of the varactors, we aim to shift the nonlinear CLS into the optical band, in line with theoretical predictions~\cite{danieli2021nonlinear, danieli2018compact}.
To achieve this, we must chirp the driver frequency up as we enter high-frequency regions due to the system's hysteresis.
In Fig.~\ref{fig:nonlinear}~(k), as the local driving frequency sweeps on the second mode of the optical band \(735^{*}\text{ kHz}\) (\(801\text{ kHz}\)), we observe the CLS transitioning to a pattern where energy oscillates between the CLS sites in an alternating (``zig-zag'') fashion and partially ``leaks'', i.e., radiates into the rest of the lattice.
This observation is consistent with simulations in panel (n), which exhibit a similar instability pattern.
We conclude that a nonlinear CLS in this electrical diamond lattice displays a special instability pattern, yet highly localized when its resonance frequency intersects the linear optical band, in contrast to Ref.~\onlinecite{danieli2018compact}, where instability destroys the CLS.

\section{Conclusions and future directions}

Using complex electrical circuits, we constructed 1D flatband lattices and observed resonant modes through local sinusoidal driving.
In a diamond lattice, we find that the driving at the flatband frequency excites a CLS.
In the stub lattice the lack of orthogonality of neighbouring CLSs precludes the observation of individual CLSs, leading instead to resonance modes with exponentially localized spatial profiles.
Finally, we found that CLSs persist in the diamond lattice when nonlinearity is introduced by replacing the capacitors with varactor diodes exhibiting symmetrical nonlinearity of capacitance, 
but that it either disintegrates entirely or becomes unstable when it is nonlinearly shifted into resonance with a dispersive band.

The clarity, as well as the qualitative and quantitative correspondence between theory, numerics and computations affirms the 
particular relevance of such linear, and, \emph{especially through our work},
nonlinear electrical lattices as a fruitful platform for exploring flat bands and CLSs, including for relevant applications such as, e.g., targeted energy  transfer~\cite{kopidakis2001}.
Naturally, numerous open questions emerge from the present work, including, e.g., whether a CLS can be distilled suitably in stub lattices, and whether more complex quasi-one-dimensional~\cite{morales2016simple}, but also two-dimensional structures~\cite{vicencio2015observation} can also be engineered.
In such settings the fate of linear and, perhaps especially, nonlinear flatband states remains an appealing and challenging task for future considerations.

\begin{acknowledgments}
    The authors acknowledge the financial support from the Institute for Basic Science (IBS) in the Republic of Korea through the project IBS-R024-D1.
    This material is based upon work supported by the US National Science Foundation under Grants DMS-2204702 and PHY-2110030 (P.G.K.).
\end{acknowledgments}

\appendix

\section{Stability of undriven nonlinear CLS}
\label{stability_of_undriven_nonlinear_cls}

\begin{figure}
    \includegraphics[width=\columnwidth]{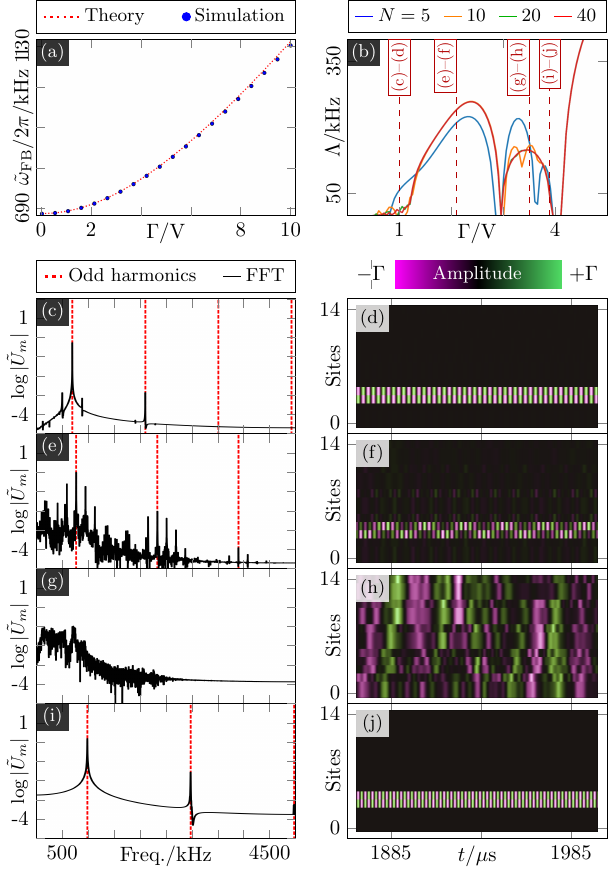}
    \caption{
        Simulation results on undriven undamped nonlinear diamond chain circuit.
        (a) Fundamental frequencies \(\tilde{\omega}_{\mathrm{FB}}/2\pi\) vs. initial CLS amplitude \(\Gamma\).
        The simulation result (lines) is compared with Eq.~\eqref{eq:nonlienar_fb_freq_approximate} (dashed red).
        (b) Computed  exponent \(\Lambda\) vs \(A\) by linear stability analysis.
        (c)-(f) Numerical results for the undriven case, with `dirty' CLS initial condition Eq.~\eqref{eq:initial_dirty_cls}, with \(\lambda = 10^{-3}\), \(N = 40\), with an initial CLS amplitude at (c)-(d) \(\Gamma = 1\) (e)-(f) \(\Gamma=2.1\), (g)-(h) \(\Gamma=2.5\), (i)-(j) \(\Gamma = 3.9\).
        The rest of the parameters of the cirucuits are the same as in the nonlinear diamond chain in the main text.
        Left panels shows the FFT result of the CLS site, \(\tilde{U}_m(\omega) = FT[U_m(t)]\).
        The right panels shows the amplitudes at all nodes after a certain time evolution (\(\sim\) 2 ms), of the first 5 unit cells among \(40\).
        The CLS is located at sites 3, 4.
    }
    \label{fig:nonlinearcomparison}
\end{figure}

In this appendix we discuss the existence and stability of a nonlinear CLS in the undriven, undamped diamond lattice in detail, using the approach similar to Ref.~\cite{danieli2018compact}.

As explained in the main text, symmetric nonlinearity is essential to find a nonlinear CLS.
The nonlinearity models we particularly consider are: quadratic, and logarithmic nonlinearity.
While both models are quadratic at low voltage, the particular choice of logarithmic modeling at high voltage provides good fits to the experimental data.

Similar to the approach in Ref.~\onlinecite{danieli2018compact}, the presence of symmetric nonlinearity preserves the exact compactneds of the CLS, \(U_{n} = -V_{n}\).
To see this, let us consider the following nonlinear CLS ansatz: \(U_n = \delta_{n,m}\Gamma f(t)\), \(V_n = -U_n\) and \(T_n = 0\).
Here \(f(t)\) is some periodic function with $f(0) = 1$, and $\Gamma$ denotes the CLS amplitude at \(t=0\).
We insert the ansatz into the nonlinear equations of motion Eq.~\eqref{eq:diamond_undriven} with the modification given in Eq.~\eqref{eq:nonlincoup}. 
For both the quadratic case and the logarithmic case at small amplitudes
\begin{align}
    \label{eq:diamond_duffing}
    \ddot{U}_m + \beta \dot{U}_m = -g_0U_m - g_1U^3_m.
\end{align}
The nonlinearity strength \(g_0 \!=\! \omega_{b0}^2(2 + \alpha)\) and \(g_1 \!=\! \omega_{b0}^2g\) characterizes the well-known Duffing oscillator.
For any solution \(U_m\), its negative partner \(V_m \!=\! -U_m\) is also a solution.
This symmetry arises from the fact that Eq.~\eqref{eq:diamond_duffing} contains only the odd powers of \(U_m\) due to the symmetric nature of the nonlinearity.
Thus, \(V_m \!=\! -U_m\) cancel out at the bottleneck sites of \(T_m\) and \(T_{m+1}\), which validates our ansatz.

As \(\Gamma\) increases, the fundamental frequency of the nonlinear CLS, denoted as \(\tilde{\omega}_{\FB}\), undergoes a shift towards higher frequencies due to the influence of the hard nonlinearity.
This effect is shown in Fig.~\ref{fig:nonlinearcomparison}~(a), which illustrates the relation between \(\Gamma\) and \(\tilde{\omega}_{\FB}\).
The dashed red line represents the estimated resonance frequency shift, obtained for the quadratic case, or logarithmic case with \(g\Gamma^2 \ll 1\), which is given by
\begin{gather}
    \label{eq:nonlienar_fb_freq_approximate}
     \tilde{\omega}_{\FB} \approx \omega_b\sqrt{(2 + 2\alpha)\left(1 + \frac{3}{4}g\Gamma^2\right)}.
\end{gather}
They can be obtained by inserting the Fourier series expanded solution into Eq.~\eqref{eq:diamond_duffing} and neglecting higher-order terms in \(g\).
Note that this approximation fails at higher voltages for the logarithmic modeling.

The stability of the nonlinear CLS becomes crucial, particularly when \(\tilde{\omega}_{\FB}\) begins to resonate with the dispersive band~\cite{danieli2018compact}, as described in Eq.~\eqref{eq:nonlienar_fb_freq_approximate}.
To investigate the stability of the nonlinear CLS, we slightly perturb it with a random vector \(\ket{\delta\psi}\),
\begin{gather}
    \label{eq:initial_dirty_cls}
    \ket{\psi(t = 0)} = \Gamma\ket{\CLS_m} + \ket{\delta \psi}, \\
    \ket{\delta \psi} = \sum_{n=0}^{N-1} \mu_n\ket{U_n} + \nu_n\ket{V_n} + \tau_n\ket{T_n},
\end{gather}
All elements \(\mu_{n},\nu_{n},\tau_{n}\) represent random noise perturbations added on the site \(U_{n}, V_{n}, T_{n}\), respectively, and uniformly distributed in \([-\lambda, \lambda]\).
The strength of the random vector is controlled by \(\lambda\) (ideally to be of infinitesimal amplitude).
The temporal dynamics of perturbed CLS over long time, as governed by Eq.~\eqref{eq:diamond_undriven} with the nonlinearity specified in Eq.~\eqref{eq:nonlincoup}, allows us to study the stability of the CLS. 
A stable CLS would be robust against any perturbations maintaining its compact localization over long time.

In Fig.~\ref{fig:nonlinearcomparison}~(c) -- (j), the spatio-temporal profiles after \(2~\mathrm{ms}\) of evolution for quadratic case (right panels) and logarithmic case (left panels),
with the initial condition Eq.~\eqref{eq:initial_dirty_cls}, with \(\lambda = 0.001\) are presented.
For \(\Gamma\leq 1\) (Fig.~\ref{fig:nonlinearcomparison}~(c) and (d)), \(\tilde{\omega}_{\FB}\) is located below the top band, the CLS remains stable, with its compactness and out-of-phase relation preserved.
For \(1 \leq \Gamma \leq 3\) (Fig.~\ref{fig:nonlinearcomparison}~(e) and (f)) we observe a quasiperiodic ``zig-zag'' instability with sharp, narrow side peaks around the harmonics in the CLS Fourier spectrum.
This ``zig-zag'' instability also appears in the driven case at certain amplitudes of driving within the top band.
This instability is of local nature, since \(\tilde{\omega}_{\FB}\) is still located below the top band.
For \(3 \leq \Gamma \leq 6.5\) the frequency  \(\tilde{\omega}_{\FB}\) resonates with the top band and its extended eigenstates, resulting in its full destruction and the excitation of extended states (Fig.~\ref{fig:nonlinearcomparison}~(g) and (h)).
We observe another narrow island of stability around \(\Gamma=3.9\) (Fig.~\ref{fig:nonlinearcomparison}~(i) and (j)), 
where the CLS regains perfect stability again (we do not currently have an explanation for that).

The result we obtain with the quadratic nonlinearity essentially agrees with the results~\cite{danieli2018compact}, in that the CLS will undergo global instability when its frequency starts to resonate with a dispersive band.
For the logarithmic nonlinearity, we still observe the three main phases of the CLS as in quadratic case: stable CLS, localized (zig-zag) instability, global instability (complete destruction of CLS), which we do not show in this paper.
However, the main difference is that, for the logarithmic case, the localized zig-zag instability is observed even when the shifted CLS frequencies are inside the top band.
This interesting difference is also observed in the experimental results in the presence of driving.
At small voltage, in Fig.~\ref{fig:nonlinear}(c) and (f), full desctruction of CLS is observed, when the resonance frequency lies in the dispersive band (green shaded area).
On the other hand, as the driver voltage increases, we observe a localized instability within the dispersive band, in Fig.~\ref{fig:nonlinear}(d)-(h).

For the quadratic case, the linear stability analysis is conducted to identify the stable and unstable regions of \(\Gamma\)~\cite{flach2004computational, flach2008discrete, flach1998discrete}.
This involves expressing the differential equation up to first order in \(\lambda\). 
The resulting linear differential equation describes the time evolution of \(\ket{\delta\psi}\).
For the quadratic case we find at the CLS site (\(n = m\))
\begin{align}
    \ddot{\mu}_{m} &= -\omega_{b0}^2 (1 \!+\! g\Gamma^2f^2(t)) \left[(2 \!+\! \alpha)\mu_{m} \!-\! \alpha \nu_{m} \!-\! \tau_{m} \!-\! \tau_{m-1} \vphantom{\sum}\right] \notag \\
    &-2\omega_{b0}^2g \Gamma^2 f^2(t)(2 \!+\! 2\alpha)\mu_{m}, \notag \\
    \ddot{\nu}_{m} &= -\omega_{b0}^2 (1 \!+\! g\Gamma^2f^2(t)) \left[(2 \!+\! \alpha)\nu_{m} \!-\! \alpha \mu_{m} \!-\! \tau_{m} \!-\! \tau_{m-1} \vphantom{\sum}\right] \notag \\
    &-2\omega_{b0}^2g\Gamma^2 f^2(t)(2 \!+\! 2\alpha)\nu_{m}.
    \label{eq:nonlinear_diamond_linearized_cls_site}
\end{align}
As before, \(\Gamma f(t)\) represents the nonlinear CLS located at \(U_m\), which is the exact solution to Eq.~\eqref{eq:diamond_duffing}.
In our analysis, we neglect higher order harmonics of $f(t)$  and consider only the first harmonic, thus focusing on the linear response.
For the non-CLS sites (\(n \neq m\)) the equation is identical to the linear case as in Eq.~\eqref{eq:diamond_undriven}.
The differential equations of \(\ket{\delta \psi}\) are expressed in a more conventional manner as follows,
\begin{align}
    \label{eq:diamond_linear_stability}
    \dot{\mathbf{y}} = \frac{d}{dt}\!
    \begin{bmatrix}
        \delta\psi \vspace{0.5em}\\ \delta\dot{\psi}
    \end{bmatrix} = 
    \begin{bmatrix}
        O & I \vspace{0.5em}\\
        H(t) & O
    \end{bmatrix}
    \begin{bmatrix}
        \delta\psi \vspace{0.5em}\\ \delta\dot{\psi}
    \end{bmatrix} = J(t)\mathbf{y},
\end{align}
where \(\delta\psi = (\cdots, \mu_{n}, \nu_{n}, \tau_{n},\cdots)^{t}\).
\(O, I\) are a \(3N \times 3N\) zero matrix and an identity matrix, respectively.
Here \(H(t)\) characterizes the Hamiltonian of the diamond lattice, specifically the right-hand side of Eq.~\eqref{eq:diamond_undriven} for a non-CLS site (\(n \neq m\)) and the right-hand side of Eq.~\eqref{eq:nonlinear_diamond_linearized_cls_site} for CLS site (\(n = m\)).
Furthermore, \(H(t)\) exhibits periodicity with a period \(T = 2\pi/\tilde{\omega}_{\FB}\) (Eq~\eqref{eq:nonlienar_fb_freq_approximate}). 
Integrating Eq.~\eqref{eq:diamond_linear_stability} yields the Floquet linear map,
\begin{align}
    \mathbf{y}(t+T) = M \mathbf{y}(t),
\end{align}
where \(M\) is the Floquet operator obtained by integrating \(J(t)\) over a period \(T\).
To study the stability of the solution, we examine the exponential growth of \(||\mathbf{y}(t)||\) over each period \(T\) in phase space.
Such an exponent is denoted as the (largest) 
exponent \(\Lambda = \ln|\overline{m}|/T\) and is measured in frequency units of \([\mathrm{kHz}]\).
Here, \(\overline{m}\) represents the largest eigenvalue of \(M\). 
An unstable CLS is characterized by \(\Lambda > 0\), while a stable CLS corresponds to \(\Lambda = 0\). 

The computation result of the 
relevant exponent is shown in Fig.~\ref{fig:nonlinearcomparison}~(b). 
At very low \(\Gamma\), \(\Lambda\) approaches zero within numerical precision, indicating that the CLS is stable at small nonlinearity.
This observation agrees with the findings presented in Fig.~\ref{fig:nonlinearcomparison}~(c) and (d).
For \(f = 1000~\mathrm{kHz}\) (\(\Gamma \approx 4.0~\mathrm{V}\)), there exists a window of stability.
Within this region, the nonlinear CLS is observed to achieve perfect stability, as illustrated in Fig.~\ref{fig:nonlinearcomparison}~(i) and (j).

Further information can be obtained by studying the eigenvectors of the Floquet operator.
In case of instability, according to bifurcation theory~\cite{blanchard2006differential}, a stable periodic orbit exists nearby.
The eigenvector corresponding to that eigenvalue \(\overline{m}\) represents the direction in phase space along which the solution changes most rapidly, towards a new stable periodic orbit.
The newly stable orbits may be a pair of asymmetric modes which therefore also cease to be compact.
The ``zig-zag'' mode (Fig.~\ref{fig:nonlinearcomparison}~(e) and (f)) is then a quasiperiodic oscillation between the two asymmetric stable modes.

\bibliographystyle{apsrev4-1}
\bibliography{flatband, mbl, general}

\end{document}